\documentclass[aps,prd,a4paper,twocolumn,amsmath,showpacs,superscriptaddress,nofootinbib,preprintnumbers]{revtex4-1}

\usepackage{verbatim}
\usepackage[T1]{fontenc}
\usepackage[utf8]{inputenc}
\usepackage[american]{babel}
\usepackage{epsfig}
\usepackage{graphicx}
\usepackage{booktabs}
\usepackage{multirow}
\usepackage{dcolumn}
\usepackage{amsmath}
\usepackage{mathtools}
\usepackage{amsfonts}
\usepackage{amssymb}
\usepackage{epstopdf}
\usepackage{bm}
\usepackage{siunitx}
\usepackage{braket}
\usepackage{enumitem}
\usepackage{soul}
\usepackage[table]{xcolor}
\usepackage{color}
\usepackage{transparent}
\usepackage{pifont}

\definecolor{navyblue}{rgb}{0.0, 0.0, 0.5}
\definecolor{royalblue}{rgb}{0.25, 0.41, 0.88}
\definecolor{cadmiumgreen}{rgb}{0.0, 0.42, 0.24}
\definecolor{blue-violet}{rgb}{0.54, 0.17, 0.89}
\definecolor{darkviolet}{rgb}{0.58, 0.0, 0.83}
\definecolor{orange(colorwheel)}{rgb}{1.0, 0.5, 0.0}

\usepackage{hyperref}
\hypersetup{
    colorlinks=true, 
    linkcolor=royalblue, 
    citecolor=magenta}

\newcommand\ee{\end{equation}}
\newcommand\be{\begin{equation}}
\newcommand\eea{\end{eqnarray}}
\newcommand\bea{\begin{eqnarray}}

\usepackage{booktabs}
\usepackage{multirow}
\usepackage{dcolumn}
\usepackage{colortbl}

\definecolor{magenta(process)}{rgb}{1.0, 0.0, 0.56}

\definecolor{darkspringgreen}{rgb}{0.09, 0.45, 0.27}

\definecolor{royalblue(web)}{rgb}{0.25, 0.41, 0.88}

\begin{document}
\title{Reconciling Planck with the local value of $H_0$ in extended parameter space}

\author{Eleonora Di Valentino}
\affiliation{Institut d’Astrophysique de Paris (UMR7095: CNRS \& UPMC-Sorbonne Universities), F-75014, Paris, France}
	
\author{Alessandro Melchiorri}
\affiliation{Physics Department and INFN, Universit\`a di Roma 
	``La Sapienza'', P.le\ Aldo Moro 2, 00185, Rome, Italy}
	
\author{Joseph Silk}
\affiliation{Institut d’Astrophysique de Paris (UMR7095: CNRS \& UPMC-Sorbonne Universities), F-75014, Paris, France}
\affiliation{AIM-Paris-Saclay, CEA/DSM/IRFU, CNRS, Univ. Paris VII, F-91191 Gif-sur-Yvette, France}
\affiliation{Department of Physics and Astronomy, The Johns Hopkins University Homewood Campus, Baltimore, MD 21218, USA }
\affiliation{BIPAC, Department of Physics, University of Oxford, Keble Road, Oxford OX1 3RH, UK}

\begin{abstract}
\noindent The recent determination of the local value of the Hubble constant by Riess et al, 2016 (hereafter R16) 
is now $3.3$ sigma higher than the value derived from the most recent CMB anisotropy data provided by 
the Planck satellite in a $\Lambda$CDM model. 
Here we perform a combined analysis of the Planck and R16 results in an extended parameter space,
varying simultaneously $12$ cosmological parameters instead of the usual $6$.
We find that a phantom-like dark energy component, with effective equation of state  $w=-1.29_{-0.12}^{+0.15}$ at $68 \%$  c.l. 
can solve the current tension between the Planck dataset and the R16 prior in an extended $\Lambda$CDM scenario.
On the other hand, the neutrino effective number is fully compatible with standard expectations.
This result is confirmed when including cosmic shear data from the CFHTLenS survey and
CMB lensing constraints from Planck. However, when BAO measurements are included we find that some of the tension with R16 remains, as also is the case when we include the supernova type Ia luminosity distances from the JLA catalog. 
\end{abstract}


\maketitle

\twocolumngrid


\section{Introduction}
\label{sec:introduction}

\noindent  Since the first data release of 2013 (\cite{planck2013}), the constraints on the Hubble constant coming from
the Planck satellite have been in significant tension with the results of Riess et al, 2011 (\cite{R11}, hereafter R11), based 
on direct measurements made with the Hubble Space Telescope. This tension was further confirmed in the 2015 Planck 
data release \cite{planck2015}. Assuming standard $\Lambda$CDM the Planck data gives $H_0=67.27\pm0.66$ km/s/Mpc
that is about two standard deviations away from the Riess et al, 2011 value of
$H_0=73.8\pm2.4$ km/s/Mpc (\cite{R11}).

Given that the Planck constraint is derived under the assumption of the "standard"
$\Lambda$CDM model, a large number of authors
(including the Planck collaboration itself, see \cite{planck2013} and \cite{planck2015}),  have proposed 
several different mechanisms to explain this tension by
considering, for example, an increased value in the effective number of
relativistic particles $N_{eff}$ (\cite{darkradiation}), phantom dark energy (see e.g. \cite{planck2013}),
interacting dark energy (\cite{interacting}), or cosmic voids (\cite{voids}). 
Cosmic variance can affect the local measurement (\cite{Ben-Dayan:2014swa}), but probably introduces too small uncertainty to explain
the discrepancy (\cite{ variance}).

On the other hand, Efstathiou (\cite{Efstathiou:2013via}) questioned the reliability of some fraction of 
the Riess et al (2011) dataset. Using the revised geometric maser distance to NGC 4258
and neglecting the Large Magellanic Cloud and Milky Way distance anchors,
Efstathiou derived a conservative constraint of $70.6 \pm 3.3$ km/s/Mpc at $68 \%$ c.l.
(EST14, hereafter), consistent in between one $\sigma$ with the Planck result.  Therefore, 
he concluded in \cite{Efstathiou:2013via} that the discrepancies between the Planck results 
and the R11 measurements  were not large enough to provide significant evidence for deviations 
from $\Lambda$CDM.

However, the recent analysis of \cite{R16} (R16, hereafter), confirmed and improved the constraint presented
in \cite{R11} with $H_0=73.24\pm1.74$ km/s/Mpc at $68 \%$ c.l., finding no compelling
argument to  not combine the three distance anchors as in \cite{Efstathiou:2013via}
and including a detailed discussion of possible systematics.
At the same time, the new constraints on the reionization optical depth, obtained with
Planck HFI data \cite{newtau}, bring the Planck constraint on $H_0$ to an even lower value,
with $H_0=66.93\pm0.62$ km/s/Mpc at $68 \%$ c.l. (see Table $8$ in \cite{newtau}).
The new R16 value, which we may refer to as  the local value of $H_0$,  is therefore more than $3.3$ standard deviations above the global value, the Planck constraint obtained assuming $\Lambda$CDM.

In other words, after three years of improved analyses and data sets, the tension in the Hubble
constant between the various  cosmological datasets not only persists but is even more 
statistically significant.

Following previous analyses (see \cite{planck2015} and {\cite{R16} and references
therein), two possible extensions to the $\Lambda$CDM scenario
have been suggested to solve the tension. It has been found that considering a neutrino
effective number $N_{eff} \sim 3.5$, i.e. the possibility of a dark radiation component,
or having a dark energy equation of state with $w \sim -1.1$ could bring the Planck
constraint into better agreement with higher values of the Hubble constant.

In this paper, we further investigate these possible solutions to the Hubble constant tension by 
performing an analysis in an extended parameter space, varying simultaneously
$12$ parameters instead of the usual $6$ assumed in $\Lambda$CDM. 
As we argued in \cite{dms2015}, many of the assumptions made
in $\Lambda$CDM are indeed not fully justified. For example, there is clearly no theoretical argument
that requires us to restrict the dark energy component to a cosmological constant.
Moreover, neutrinos are massive and there is no current laboratory measurement that 
could constraint their absolute mass scale to be less than, say,  
$\Sigma m_{\nu} < 1 eV$. Assuming the minimal value of $\Sigma m_{\nu} =0.06$ eV 
as in $\Lambda$CDM could therefore introduce a strong bias in the analysis since 
it is equivalent to removing a large portion of the physically allowed parameter space.
Hence especially in view of the new precise measurements
made by Planck, it seems reasonable to consider a larger parameter space.

It is also important to stress that simply increasing the number of parameters would
not necessarily bring the two datasets in agreement. The neutrino mass,
for example, anti-correlates with the value of the Hubble constant when
constrained from CMB data, and the Planck constraint would be even
lower when variations in $\Sigma m_{\nu}$ are considered.

Following the method presented in \cite{dms2015}, we therefore consider as additional 
parameters the dark energy equation of state $w$, the neutrino effective number
$N_{eff}$ , the running of the spectral index $d n_s / dlnk$, the tensor to scalar ratio $r$,
the neutrino mass $\Sigma m_{\nu}$ and, finally, the amplitude of the gravitational
lensing on the CMB angular spectra $A_{lens}$ (see \cite{Calabrese:2008rt} for
a definition).  The inclusion of the last parameter comes from the Planck data itself that suggests
an anomalous value of $A_{lens}=1.15^{+0.13}_{-0.12}$ at $95 \%$ c.l. \cite{newtau}, but see also \cite{Grandis:2016fwl}. 

However, respect to \cite{dms2015}, here we include the new R16 result, studying the compatibility
not only with the Planck data, but also with several combination of datasets.
Indeed the goal of this paper is to identify a new "concordance" model in an extended parameter
space, where the new R16 result could be accommodated.

Moreover, another anomaly is present when the Planck dataset alone si considered:
indeed, Planck is suggesting also a non flat universe, with positive curvature
such that the curvature density parameter is constrained to be 
$\Omega_k=-0.052_{-0.055}^{+0.049}$ at $95 \%$ c.l.. (see \cite{planck2015}). It is therefore interesting
to consider also this possibility and in this paper we further extend the analysis
presented in  \cite{dms2015} by considering an extended parameter space where
curvature, instead of $A_{lens}$, is varied.

Our brief paper is structured as follows: in the next Section we describe the data analysis 
method adopted, in Section III we present our results and in Section IV we derive
our conclusions.

\section{Method}
\label{sec:method}

As in \cite{dms2015} we analyze current cosmological data by making use of
publicly available code \texttt{cosmomc}~\cite{Lewis:2002ah, Lewis:2013hha}.

As discussed in the introduction, following \cite{dms2015}, 
we consider an extended $\Lambda$CDM scenario where we vary a total
of $12$ cosmological parameters simultaneously.

We indeed vary the "standard" six parameters of the $\Lambda$CDM model:
the baryon $\omega_b$ and cold dark matter $\omega_c$ energy densities,
the angular diameter distance to the sound horizon at last scattering $\theta$,
the amplitude $A_s$ and tilt $n_s$ of primordial scalar fluctuations and the
reionization optical depth $\tau$.
In addiction to these parameters, we vary at the same time also 
 $6$ extra parameters: the absolute neutrino mass scale
$\Sigma m_{\nu}$, the neutrino effective number $N_{eff}$,
the tensor-to-scalar ratio $r$, the running of the scalar
spectral index $d n_s / dln k$, the dark energy equation of state 
$w$ and the lensing amplitude in temperature and polarization
angular spectra $A_{lens}$.

Moreover, as mentioned in the introduction, we also consider a slightly different extended 
parameter space by fixing the values of the neutrino effective number and of the lensing
amplitude to their LCDM values of $N_{eff}=3.046$ and $A_{lens}=1$, but
letting now the curvature density $\Omega_k$ to vary.
In this way we could not only test the possibility of a curved universe, as suggested by 
Planck data alone, but also in someway quantify how much the results could
depend on the variation of $A_{lens}$ that is indeed an effective 
parameter with an unclear origin.

Our main dataset consists of CMB  temperature and polarization anisotropies
from the Planck 2015 data release (\cite{Aghanim:2015xee}). 
In what follows, we refer to this dataset simply as ``\emph{Planck}''. 

Together with the R16 constraint on the Hubble constant, that we
treat as an external gaussian prior of $H_0=73.20\pm1.74$ km/s/Mpc at $68 \%$ c.l.,
we also consider the following additional datasets:

\begin{itemize}

\item The collection of Baryonic Acoustic Observations (BAO)
(6dFGS \cite{Beutler:2011hx}, SDSS-MGS \cite{Ross:2014qpa}, BOSS LOWZ \cite{Anderson:2013zyy} 
and CMASS-DR11 \cite{Anderson:2013zyy} \emph{BAO};
\item The luminosity distances of supernovae type Ia from the Joint Light-curve Analysis catalog (JLA) \cite{betoule};
\item \emph{Planck} measurements of the CMB \emph{lensing} potential power spectrum $C^{\phi\phi}_\ell$ \cite{Ade:2015zua};
\item weak lensing (WL) data from the CFHTLenS survey \cite{Heymans:2012gg, Erben:2012zw}, taking wavenumbers 
$k\leq 1.5 h\,\mathrm{Mpc}^{-1}$\cite{planck2015, Kitching:2014dtq};

\end{itemize}

\section{Results}
\label{sec:results}
\begin{table*}
\begin{center}\footnotesize
\scalebox{1.04}{\begin{tabular}{lccccccc}
\hline \hline
         & Planck & Planck  & Planck & Planck &Planck&Planck\\                     
         &      &        + R16  & +R16+BAO       &    +R16+JLA &+R16+ WL&+R16+lensing\\  
\hline
\hspace{1mm}\\

$\Omega_{\textrm{b}}h^2$& $0.02239\,\pm 0.00030 $& $0.02239\,\pm 0.00029$    & $0.02258\,_{-0.00032}^{+0.00026} $& $0.02270\,\pm 0.00025$& $0.02253\,\pm 0.00029$ & $0.02214\,\pm 0.00027$   \\
\hspace{1mm}\\

$\Omega_{\textrm{c}}h^2$& $0.1186\,\pm 0.0035$& $0.1187\,\pm 0.0036$    & $0.1209\,_{-0.0036}^{+0.0032} $& $0.1218\,\pm 0.0034$ & $0.1188\,\pm0.0036$ & $0.1176\,\pm 0.0035$   \\
\hspace{1mm}\\

$\tau$& $0.058\,\pm 0.021$& $0.058\,_{-0.023}^{+0.021}$    & $0.058\,\pm 0.021$& $0.059\,\pm 0.021$& $0.050\,_{-0.022}^{+0.019}$ & $0.058\,\pm 0.021$    \\
\hspace{1mm}\\

$n_S$& $0.967\,\pm 0.013$& $0.967\,\pm 0.013$    & $0.976\,\pm 0.12 $& $0.981\,\pm 0.011$& $0.973\,\pm 0.012$ & $0.959\,\pm 0.012$   \\
\hspace{1mm}\\

$log(10^{10}A_S)$& $3.048\,\pm 0.043$& $3.048\,_{-0.048}^{+0.043}$    & $3.053\,\pm 0.043$& $3.056\,\pm 0.043$& $3.030\,\pm 0.041$ & $3.043\,\pm 0.043$    \\
\hspace{1mm}\\

$H_0$ &      $>67.1$&      $ 73.5\,\pm 1.9$ & $ 71.3\,\pm 1.6$   &  $ 70.9\,\pm 1.5$ &  $ 73.6\,\pm 1.9$  &  $ 73.7\,\pm 2.0$ \\
\hspace{1mm}\\

$\sigma_8$   & $ 0.81\,_{-0.12}^{+0.16}$   & $ 0.804\,_{-0.044}^{+0.056}$   & $ 0.788\,\pm 0.036$ &  $ 0.785\,_{-0.037}^{+0.056}$ &  $ 0.786\,_{-0.042}^{+0.053}$ &  $ 0.827\,\pm0.039$  \\
\hspace{1mm}\\

$\sum m_{\nu}$ [eV] &      $<0.53$&      $<0.512$ & $0.35\, _{-0.25}^{+0.16}$   & $<0.384$ & $0.43\, _{-0.41}^{+0.12}$ & $0.32\, _{-0.24}^{+0.14}$  \\
\hspace{1mm}\\

$w$ &  $-1.32\,_{-0.67}^{+0.47}$ &  $-1.29\,_{-0.12}^{+0.15}$  & $-1.14\,_{-0.10}^{+0.12}$&  $-1.079\,_{-0.057}^{+0.072}$ &  $-1.25\,_{-0.11}^{+0.13}$ &  $-1.33\,_{-0.12}^{+0.15}$   \\
\hspace{1mm}\\

$N_{\rm eff}$ &  $3.08\,_{-0.30}^{+0.26}$ &  $3.09\,_{-0.31}^{+0.26}$  & $3.26\,_{-0.28}^{+0.24}$&  $3.37\,_{-0.28}^{+0.24}$ &  $3.17\,_{-0.31}^{+0.26}$ &  $2.94\,\pm0.25$  \\
\hspace{1mm}\\

$A_{\rm lens}$ &  $1.21\,_{-0.14}^{+0.09}$ &  $1.18\,_{-0.11}^{+0.09}$  & $1.210\,\pm 0.095$&  $1.22\,_{-0.11}^{+0.09}$ &  $1.233\,_{-0.099}^{+0.085}$ &  $1.031\pm0.062$   \\
\hspace{1mm}\\

$\frac{dn_S}{d\, ln\, k}$ &  $-0.0034\,\pm 0.0098$ &  $-0.003\,_{-0.011}^{+0.010}$  & $-0.0003\,\pm 0.0091$&  $0.001\,_{-0.011}^{+0.009}$ &  $-0.0003\,\pm0.0097$ &  $-0.0054\,\pm0.0090$   \\
\hspace{1mm}\\

$r$ &  $<0.0911$ &  $<0.0934$  & $<0.0974$&  $<0.0943$ &  $<0.099$& $<0.0856$   \\
\hspace{1mm}\\
\hline
\hline

\end{tabular}}
\caption{$68 \% $ c.l. constraints on cosmological parameters in our extended $12$ parameters scenario from different combinations 
of datasets.}
\label{table}
\end{center}
\end{table*}

\begin{figure}[!hbt]
\includegraphics[width=0.48\textwidth]{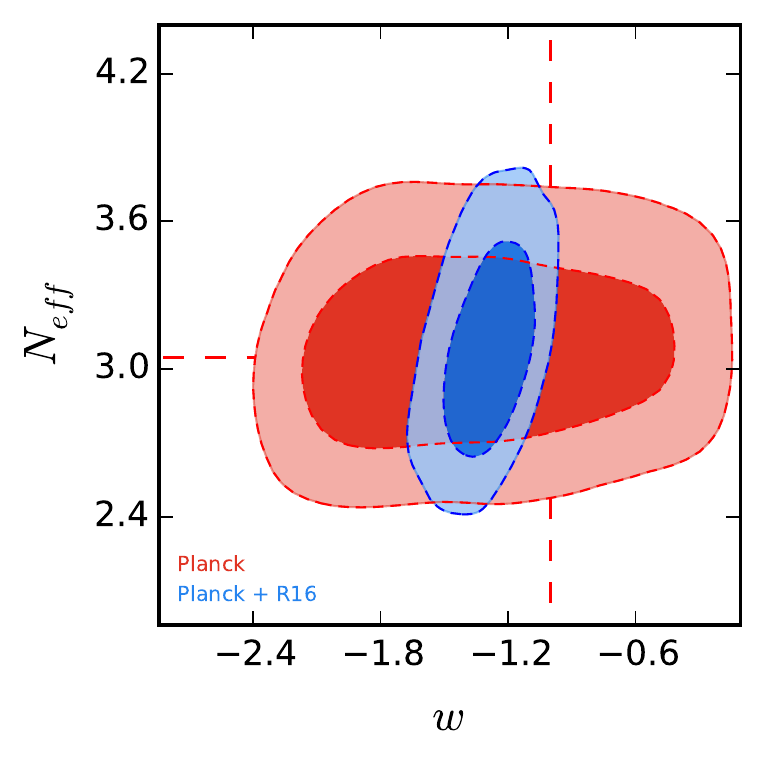}
\caption{\footnotesize{Constraints at $68 \%$ and $95 \%$ c.l. on the $N_{eff}$ vs $w$ plane assuming the Planck data set
with and without the R16 prior on $H_0$. As one can see, when the R16 prior is included, a preference for $w<-1$ is clearly present,
while $N_{eff}$ is consistent with the standard expectations. An extended $\Lambda$CDM theoretical framework of $12$ parameters is assumed in the analysis.}}
\label{figure}
\end{figure}

Our main results are reported in Table \ref{table} where we report the constraints at $68 \% $ c.l. on the
$12$ parameters of our extended scenario.
As discussed in the previous section, we consider the Planck dataset (temperature and polarization)
plus the new R16 prior in combination with BAO, JLA, CFHTLenS and Planck CMB lensing data sets. 
For comparison, we also consider the Planck data set alone.

We found that the Planck+R16 data set provides a reasonable increase in the effective chi-square value  
of $\Delta \chi_{eff}^2 \sim 0.9$ with respect to the Planck data set alone, with one single additional data point.
In other words, the R16 prior is fully compatible with the Planck data in our extended $\Lambda$CDM scenario.
It is therefore interesting to understand which of the extra parameters contributes to restoring
the agreement between Planck and R16. By looking at the extra parameters, we  notice that while
 the neutrino effective number is compatible with its standard value of $N_{eff}=3.046$, 
the dark energy equation of state is below $-1$ at the level of $\sim 2$ sigma, hinting at  new physics in
the dark energy sector. We also see that the $A_{L}$ is larger than its standard value at more 
than $2$ standard deviations. However this anomaly is driven by the Planck dataset and the inclusion of the R16 prior
does not significantly affect its statistical significance.

When the BAO dataset is included, the indication for $w<-1$ is much less significant. 
Moreover, we note an increase in the value of $N_{eff}$, even if is still in agreement
with its standard value. More interestingly, the inclusion of the BAO dataset shows
an indication at one sigma for a neutrino mass with $\Sigma m_{\nu}=0.35\, _{-0.25}^{+0.16}$ at $68 \%$ c.l..
The inclusion of the R16 prior in the Planck+BAO dataset increases the 
effective chi-square by  $\Delta \chi^2 \sim 4.5$,
suggesting a  tension between the R16 prior and Planck+BAO even in
a $12$ parameter extension. This is clearly driven by the value of the Hubble constant
from the Planck+BAO dataset that is constrained to be 
$H_0=68.4_{-4.1}^{+4.3}$ at $95\%$ c.l. \cite{dms2015}, i.e. lower than
the R16 prior. 
 
The inclusion of the JLA dataset, on the other hand, suggests at about one standard deviation
a value for $N_{eff} > 3.046$ and a dark equation of state $w<-1$.
In this case, the effective chi-square value  when a R16 prior is included in a Planck+JLA
analysis increases by $\Delta \chi^2 \sim 4.1$, indicating, as in the case of BAO, 
a tension between the Planck+JLA dataset and the R16 prior.

Vice  versa, when the WL and CMB lensing datasets are included, we have 
again an indication for $w<-1$ (at $1.7$ sigma for WL and $2.4$ sigma
for CMB lensing) while the $\chi^2$ is not significantly affected by
the inclusion of the R16 prior. We indeed found an increase in the effective
chi-square of $\Delta \chi^2 \sim 0.8$
when the R16 prior is included in the analysis of the Planck+WL dataset and
$\Delta \chi^2 \sim 1$ when it is included in the analysis of the 
Planck+lensing dataset.

\begin{table*}
\begin{center}\footnotesize

\scalebox{1.04}{\begin{tabular}{lcccc}
\hline \hline
         & Planck & Planck  & Planck\\                     
         &             &+ R16+lensing  & +R16+BAO\\  
\hline
\hspace{1mm}\\

$\Omega_{\textrm{b}}h^2$& $0.02238\,\pm 0.00018 $& $0.02221\,\pm 0.00018$    & $0.02232\,_{-0.00018}^{+0.00019} $\\
\hspace{1mm}\\
$\Omega_{\textrm{c}}h^2$& $0.1183\,\pm 0.0016$& $0.1191\,\pm 0.0015$    & $0.1195\,_{-0.0015}^{+0.0015} $\\
\hspace{1mm}\\

$\tau$& $0.054\,\pm 0.021$& $0.056\,_{-0.020}^{+0.021}$    & $0.083\,\pm 0.019$\\
\hspace{1mm}\\

$n_S$& $0.9675\,\pm 0.0055$& $0.9641\,\pm 0.0055$    & $0.9646\,\pm 0.0053 $\\
\hspace{1mm}\\

$log(10^{10}A_S)$& $3.039\,\pm 0.042$& $3.043\,_{-0.041}^{+0.042}$    & $3.101\,\pm 0.037$\\

\hspace{1mm}\\

$H_0$ &      $51_{-10}^{+6}$&      $ 73.7\,\pm 2.0$ & $ 71.3\,\pm 1.6$ \\
\hspace{1mm}\\

$\sigma_8$   & $ 0.724\,_{-0.13}^{+0.063}$   & $ 0.845\,_{-0.026}^{+0.026}$   & $ 0.861^{+0.036}_{-0.025}$\\
\hspace{1mm}\\

$\sum m_{\nu}$ [eV] &      $0.29_{-0.24}^{+0.13}$&      $0.32\pm0.16$ & $<0.172$\\
\hspace{1mm}\\

$w$ &  $-0.99\,_{-0.45}^{+0.72}$ &  $-1.45\,_{-0.19}^{+0.25}$  & $-1.193\,_{-0.10}^{+0.088}$\\
\hspace{1mm}\\

$\Omega_k$ &  $-0.067^{+0.053}_{-0.025}$ &  $-0.0046^{+0.0053}_{-0.0064}$  & $-0.0018_{-0.0034}^{+0.0026}$\\
\hspace{1mm}\\

$\frac{dn_S}{d\, ln\, k}$ &  $-0.0022\,\pm 0.0074$ &  $-0.0019\,_{-0.0077}^{+0.0078}$  & $-0.0073\,\pm 0.0076$\\
\hspace{1mm}\\
$r$ &  $<0.0977$ &  $<0.0834$  & $<0.0722$\\
\hspace{1mm}\\
\hline
\hline
\end{tabular}}
\caption{$68 \% $ c.l. constraints on cosmological parameters in our extended $11$ parameters scenario that includes variations
in $\Omega_k$ from different combinations  of datasets.}
\label{table}
\end{center}
\end{table*}

In order to further test the stability of our results under a different choice of the parameter space, we have also considered
the possibility of a "less extended" parameter space of $11$ parameters. In this case, we fix the neutrino effective number and the
lensing amplitude to their LCDM values of $N_{eff}=3.046$ and $A_{lens}=1$, but letting this time the curvature parameter
$\Omega_k$ to vary freely.
Our results are reported in Table 2. As we can see from the first column, in this parameter space the Hubble constant 
is constrained from Planck to be $H_0=51_{-10}^{+6}$ at $68 \%$ c.l.. The Planck dataset alone is therefore not compatible anymore
with the R16 prior despite the significant increase in the parameter space. Indeed, while the effect of introducing variations in the
neutrino number $N_{eff}$ and the lensing amplitude $A_L$ is to allow a better compatibility of larger values of $H_0$, the introduction of curvature produces exactly the opposite effect. We can therefore claim that a positive curvature, as suggested by Planck data alone, does not solve the tension between Planck and R16 on the value of the Hubble parameter, even in a $11$ parameters space.
It is interesting to study the compatibility with R16 when additional datasets as BAO or lensing are included, since their main effect, as
discussed in \cite{planck2015}, is to constrain curvature to be very close to zero.
We have indeed found (always in this new $11$ parameters space) that a Planck+BAO or Planck+Lensing analysis constrain the Hubble constant to $H_0=73.7\pm2.0$ km/s/Mpc and $H_0=67_{-20}^{+10}$ km/s/Mpc respectively, at $68 \%$ c.l., i.e. to values that are now in agreement with the R16 prior.
We report in the second and third columns of Table 2 the constraints on the $11$ cosmological parameters for the Planck+lensing+R16 and Planck+BAO+R16 datasets.
We can notice that in both cases the curvature is always extremely close to zero and in both cases the 
equation of state $w$ is below $-1$ at about $95 \%$ c.l.. In the Planck+lensing+R16 case we have an indication
at about $95 \%$ c.l. for a neutrino mass, while the optical depth is significantly larger for Planck+BAO+R16.
We can therefore conclude that when restricted to a $11$ parameters space and after fixing the curvature anomaly
including the BAO or lensing dataset, we found that the combined datasets suggest, again, $w<-1$ at about $95 \%$ c.l..

\section{Conclusions}
\label{sec:conclusions}

The recent determination of the local value of the Hubble constant by R16 
is now $3.3$ sigma higher than the value determined by measurements of CMB anisotropies made by 
the Planck satellite mission in a $\Lambda$CDM model.
While the presence of systematics is not yet excluded, it is interesting to investigate what kind of new physics
could solve the discrepancy.  In this brief paper,
we have performed a combined analysis of the Planck and R16 result in an extended parameter space,
varying simultaneously $12$ cosmological parameters instead of the usual $6$ of $\Lambda$CDM, 
since in this scenario a higher value of $H_0$ is naturally allowed. 
We found that in this $12$ parameter space, the tension is reduced with  $N_{eff}=3.09_{-0.31}^{+0.26}$ at $68 \%$ c.l.,
in very good agreement with the standard expectations, $H_0=73.5 \pm 2.9$ km/s/Mpc at $68 \%$ c.l., 
and $w=-1.29^{+0.15}_{-0.12}$, suggesting  a phantom-like dark energy component at the level of $2$ sigma. 
Moreover, this extended scenario prefers a lower value of the reionization optical depth $\tau=0.058\pm0.021$, in complete agreement with the new value provided by Planck HFI data \cite{newtau}.
This result and the indication for $w<-1$ are confirmed when cosmic shear data from the CFHTLenS survey or CMB lensing data
from the Planck maps are included in the analysis.
However, when BAO measurements are included we get $N_{eff}=3.26_{-0.28}^{+0.24}$ at $68 \%$ c.l., 
$H_0=71.3 \pm 1.6$ km/s/Mpc at $68 \%$ c.l., and $w=-1.14^{+0.12}_{-0.10}$,
with the indication for $w<-1$ now present at just $\sim 1.1$ sigma.
The inclusion of the R16 prior in the Planck+BAO dataset produces a worse fit of $\Delta \chi^2 \sim 4.5$. This is due to the tension at the level $1.7$ sigma existing between the $H_0$ value provided by Planck + BAO; also in this extended 12 parameter space ($H_0=68.4_{-4.1}^{+4.3}$ at $95\%$ c.l. \cite{dms2015}), and R16.

Including the supernova type Ia luminosity distances from the JLA catalog gives
 $N_{eff}=3.37_{-0.28}^{+0.24}$ at $68 \%$ c.l.,  $H_0=70.9 \pm 1.5$ km/s/Mpc at $68 \%$ c.l., 
 and $w=-1.079^{+0.072}_{-0.057}$, showing non-standard values for {\it both} $w$ and $N_{eff}$
 at one sigma level. The chi-square value of the best fit increases by $\Delta \chi^2 \sim 4.1$ when
 a R16 prior is included in a Planck+JLA analysis, again due to a tension existing between the datasets.
In fact, Planck + JLA prefers $H_0=67.4_{-4.2}^{+4.4}$ at $95\%$ c.l. \cite{dms2015} in this extended scenario, 
almost two sigma lower with respect to R16.

Finally, we have considered a new, slightly different, extended parameter space letting
curvature to vary. While curvature does not solve the tension between Planck and R16 on the
Hubble constant, we have found that a combination of datasets as Planck+BAO and
Planck+lensing can be put in agreement with the R16 prior by letting, once again, the
equation of state $w$ to be $<-1$.

We can therefore conclude that a variation in $w$ can solve the current tension between
the Planck dataset and the R16 prior in an extended $\Lambda$CDM scenario and that
this result is confirmed when including the WL and CMB lensing datasets.
Clearly, this indication for $w<-1$ could hide a more complicated dark energy model.
Indeed, since we assumed $w$ as constant with time, this can smear out information about 
$w$ and its time variation (see e.g. \cite{weff}). Apart from phantom dark energy models
with a genuine equation of state $w<-1$ (see e.g. \cite{phantom}),   models with a time-varying 
equation of state as interacting dark energy  could also provide an effective value 
(averaged over redshift) of $w_{eff}<-1$  as  obtained here (\cite{das},\cite{Cai:2009zp}).
Interestingly, modified gravity models, such as, for example,
the Hu and Sawicky model \cite{husawicki},  could also provide a value for $w_{eff}<-1$.
Modified gravity could also account for the $A_{lens}$ anomaly
(see e.g. \cite{dmsmg}).

However the tension with the R16 value persists when the Planck+BAO or Planck+JLA datasets 
are considered, suggesting an even more complicated extension might be needed for $\Lambda$CDM, or, maybe more likely,  
systematic errors between the data sets. Since the increase in the number of parameters considered here 
is already significant, the presence of systematics in the datasets provides, in our opinion,
a more conservative explanation. However, even if not all the datasets considered point in this direction,
most of them indicates that the  LCDM model may still  be incorrect and several tensions are solved by introducing
new physics. Future data from CMB experiments and galaxy surveys 
as DESI and EUCLID will certainly clarify the issue.

\section*{Acknowledgements}

\noindent We thank Martin White for useful comments. AM is supported by the research grant Theoretical Astroparticle Physics number 2012CPPYP7 under the program PRIN 2012 funded by MIUR and by TASP, iniziativa specifica INFN.  This work has been done within the Labex ILP (reference ANR-10-LABX-63) part of the Idex SUPER, and received financial state aid managed by the Agence Nationale de la Recherche, as part of the programme Investissements d'avenir under the reference ANR-11-IDEX-0004-02. EDV acknowledges the support of the European Research Council via the Grant  number 267117 (DARK, P.I. Joseph Silk).


\end{document}